\documentclass[preprint,amsmath,amssymb]{revtex4-1}
\usepackage{graphicx, color}
\usepackage{caption}
\usepackage{subcaption}

\newcommand{\mboxss}[1]{\mbox{\scriptsize #1}}
\let\oldAA\AA
\renewcommand{\AA}{\text{\normalfont\oldAA}}
\newcommand{\eps}{\mbox{$\epsilon$}}
\newcommand{\kboltz}[1]{\mbox{$k_{\mbox{\scriptsize B}}^{#1}$}}

\newcommand{\lrangSUscr}[3]{\mbox{$\left\langle #1 \right\rangle_{\mboxss{#2}}^{\mboxss{#3}}$}}

\begin{document}


\title{Implementing and constraining higher fidelity kinetics for DPAL models} 



\author{H. J. Cambier}
\email[]{hal.cambier.ctr@us.af.mil}

\author{T. J. Madden}
\affiliation{Kirtland AFRL}


\date{\today}

\begin{abstract}
   Ionization, hydrocarbon breakdown, and other exotic processes can harm diode-pumped alkali laser (DPAL) performance and components.  We develop a physical picture of these processes, including those that drive a non-Maxwell-Boltzmann distribution of electrons, and describe an efficient approach to solve these kinetics while resolving trace species, and enforcing conservation laws.  Comparing the model to time-dependent experiments suggests that recombination and supporting processes are weaker than na\"{i}vely expected under relevant conditions, while methane seems to improve performance in the lab more than it does in the model.  Overall, this work highlights the importance of tracking the true electron energy distribution, and how incisive experiments with time-dependent driving are.  We also use the model to emphasize how ionization may pose more immediate heat loading problems in devices.
\end{abstract}


\maketitle 


\section{Introduction}
\label{sec:GenIntro}

   {For decades, the kinetics of alkali vapors have garnered interest given the role they play in atomic physics experiments, atomic line filters, thermionic generators, etc.  More recently, mid-infrared diodes have become spectrally narrow enough to excite individual fine-structure components of the first resonance level in alkali like potassium, rubidium, and cesium.  This allowed for a fine-structure-specific three-level lasing scheme in buffered alkali vapors, and introduced a new application: the diode-pumped alkali laser (DPAL).

   Past research into optically-pumped alkali vapors outside the DPAL context has shown it is relatively easy for the alkali atoms to reach higher excitation levels, ionize, or undergo {chemical} reactions, {causing concerns for DPALs as well.}}  {In earlier work,} Wu\cite{2009_Wu_Thesis} presented {a} model of DPAL ionization including laser rate equations, {but assumed that the electrons were in a Maxwell-Boltzmann distribution with the same temperature as the buffer gas.}  Oliker et al.\cite{2014_OlikerEt_LossProcesses} studied {the effects of} similar kinetics in a model {that included computational fluid dynamics, and ray-tracing.  Wallerestein et al.\cite{2018_WallersteinPerramRice} considered an even broader set of processes involving the bound electrons, and emphasized building intuition for the hierarchy of processes based on their timescales, and likely availability of reactants.}

   {However,} Zatsarinny et al.\cite{2014_ZatsarinnyEtc_CsSigmas}, and Markosyan et al.\cite{2016_MarkosyanKushner_16CsDpalPlasma,2018_Markosyan_MoreDpalPlasma}, {did} consider a non-Maxwell-Boltzmann (non-MB) electron energy distribution through the use of finely-grained lookup tables\cite{2004_StaffordKushner}.  We also relax the Maxwell-Boltzmann assumption, but model, and evolve the electron distribution a bit differently, and focus on experimental comparisons.   {Because processes like electron impact transitions (EIT), electron impact ionization (EII), and recombination are sensitive to electron energy, {relaxing assumptions about the energy distribution} avoids underestimating ionization, and its effects.  This {generalization} also poses numerical issues like stiffness, and maintaining positivity, which have been tackled for decades as well.}  Some relevant techniques see little use in this context though, so we discuss the implementation of one approach that proved instrumental{.  For the readers' benefit, we also mention issues with simple implementations of other approaches.}

   {The paper is organized as follows.}  We give an overview of major processes, including those affecting the electron energy distribution in Section \ref{sec:PhysIntro}, while relegating details on inputs to Section \ref{sec:A_FIRST}-\ref{sec:A_LAST}.  We describe the numerical implementation in Section \ref{sec:NumIntro}.  In Section \ref{sec:Applying}, we compare the model to experiments by Zhdanov et al.\cite{2017_ZhdanovEt_KDPAL_BufferStudy}.  {Here, we also demonstrate, and explain the tendency for non-MB electrons to increase heat loading.}  We present our conclusions in Section \ref{sec:Conc}.

\section{Developing the physical picture}
\label{sec:PhysIntro}

\begin{figure}[!htbp]
      \includegraphics[width=0.6\textwidth]{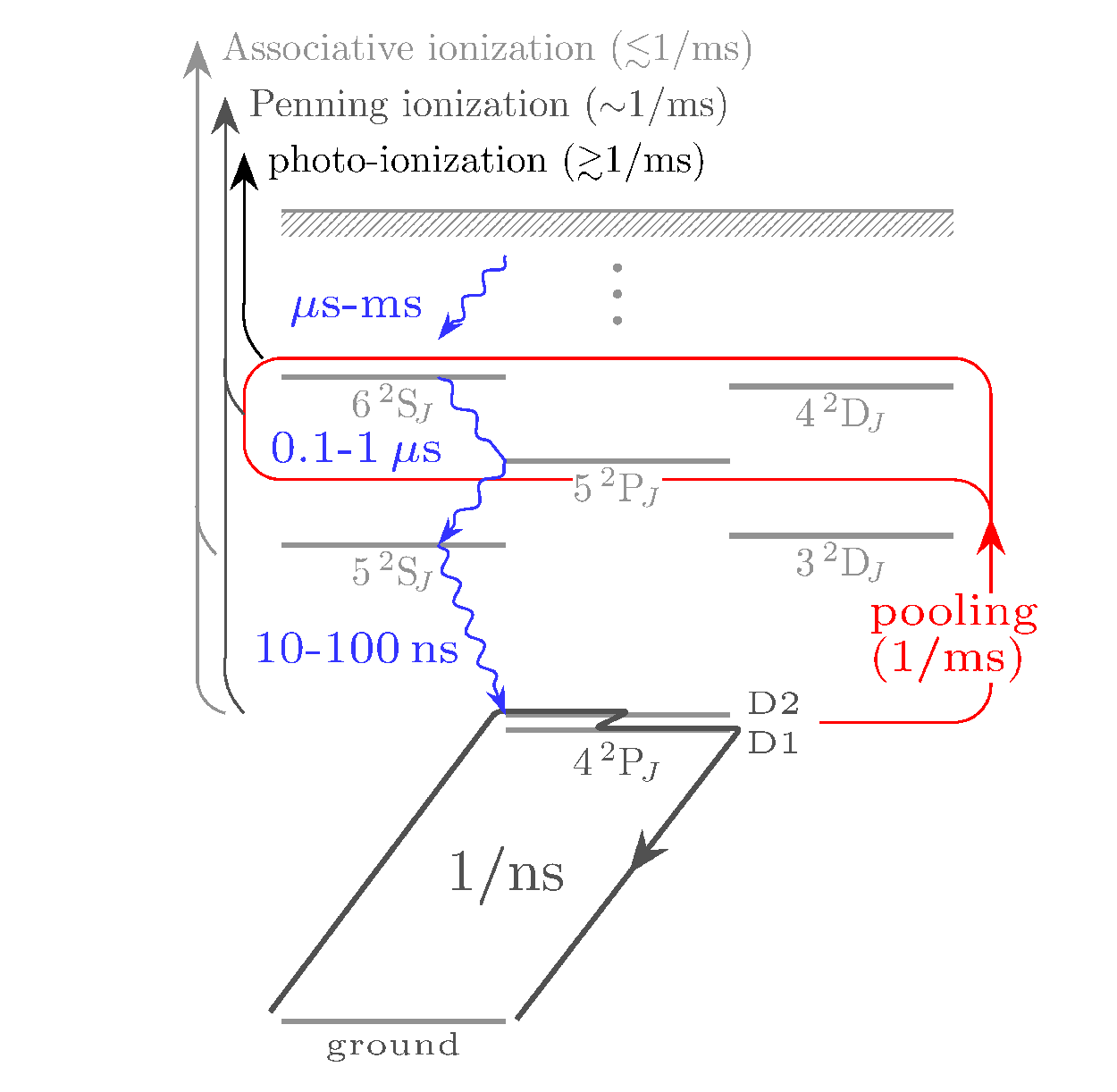}
      \caption{\label{fig:CartoonK}
      Schematic potassium Grotrian diagram showing fine-structure multiplets modeled, and ideal lasing scheme, energy pooling among resonance-level atoms, and dominant ionization channels (left), along with their characterstic rates.  Radiative decays with characteristic timescales, are also shown.
      }
   \end{figure}

   {A typical DPAL medium includes around 0.9 atmospheres of helium, 0.1 atmospheres of some hydrocarbon buffer, and 1-10 parts per million of alkali vapor (e.g. 10$^{13}$-10$^{14}$/cm$^3$), at temperatures around 400-500K.}  Diodes pump alkali atoms to the upper, more degenerate, resonance sub-level, ``D2'', (4\,$^2$P$_{3/2}$ for K).  Fine-structure mixing collisions with the buffer drive transitions to the lower resonance sub-level, ``D1'' (4\,$^2$P$_{1/2}$), which is the state used for lasing (fig.~\ref{fig:CartoonK}).  Higher pressure designs may achieve enough mixing with less or no molecular buffer, thus avoiding thermal-lensing, and buffer chemistry issues.  {For example,} Zhdanov et al. \cite{2007_ZhdanovKnize_KHeDemo} showed that just two atmospheres of helium sufficed to mix {D1 and D2} in potassium.

   Deviation from ideal operation begins when energy-pooling collisions {(EP)} between resonant-level alkali populate ``pioneer'' Rydberg levels with rates\cite{1997_NamiotkaHA_EP_K,1983_BarbierCheret_EP_Rb,1996_VadlaEtc_CsPooling,1996_JabbourEtal_CsEPdata} spanning $10^{-11}$ to $10^{-9}\text{cm$^3$/s}$.  Collisional ionization (CI), (single) photo-ionization (PhI), {are the fastest deleterious pathways out of these levels.  Decays (unhindered by radiation trapping) to the 5\,$^2$S$_J$ and 3\,$^2$D$_J$ multiplets, and energy-quenching collisions (EQ)} will counter these processes{}.  Buffer collisions dominate EQ, and Earl and Herm \cite{1974J_EarlHerm_EQ_NaKxMs} find that methane quenches the 5\,$^2$P$_J$ level with a roughly 60\AA$^{2}$ cross section.  For our default model, we assume this 5\,$^2$P$_J$ quenching populates 3\,$^2$D$_J$ and 5\,$^2$S$_J$, and that methane quenches 4\,$^2$D$_J$ and 6\,$^2$S$_J$ to 5\,$^2$P$_J$ with a similar cross section given the similarity of all energy gaps involved (see Section \ref{sec:A_Heavies}).  Helium quenching becomes more significant at much higher levels, which we do not resolve in this work.  {The following subsections discuss some processes, or aspects of them, that tend to receive less attention, while the appendices give more specific implementation details.}

   \subsection{Collisional and photon ionization}
   Explicit data for potassium is scarce, but data on other alkali, and the intermolecular potential curves\cite{1988_BrencherEtal_CI_K} indicate that Penning ionization involving a resonance-level, and a pioneer-level atom -- producing just an atomic ion -- forms the primary collisional ionization (CI) channel.  {Looking at rubidium, Barbier et al.\cite{1987_BarbierPC_CI_RbAnalogExp} found that associative ionization between one resonance-level, and one higher energy alkali atom becomes orders of magnitude weaker than Penning ionization}, and they were able to explain this theoretically by invoking electronic exchange\cite{1987_BarbierPC_CI_RbAnalogy}.  This leaves collisions between one D1/D2, and one 5\,$^2$S$_J$/3\,$^2$D$_J$ atom as the most plausible route for direct ${\rm K}_2^+$ production, but processes described below can influence the dimer cation fraction far more effectively.

   {Predictions by Aymar et al. \cite{1976_AymarEt_yIs_Kspd}, and Zatsarinny et al. \cite{2010_ZatsarinnyTayal_yIs_Kspd} for photo-ionization (PhI) cross-sections near D1 and D2 wavelengths} imply that photo-ionization exceeds any CI channel above about 10kW/cm$^2$ for $10^{14}$/cm$^3$ alkali densities.  Note that both the dominant CI, and PhI channels give {free electrons an initial kinetic energy of 0.3-0.6eV.}

   \begin{figure}[!htbp]
      \includegraphics[width=0.5\textwidth]{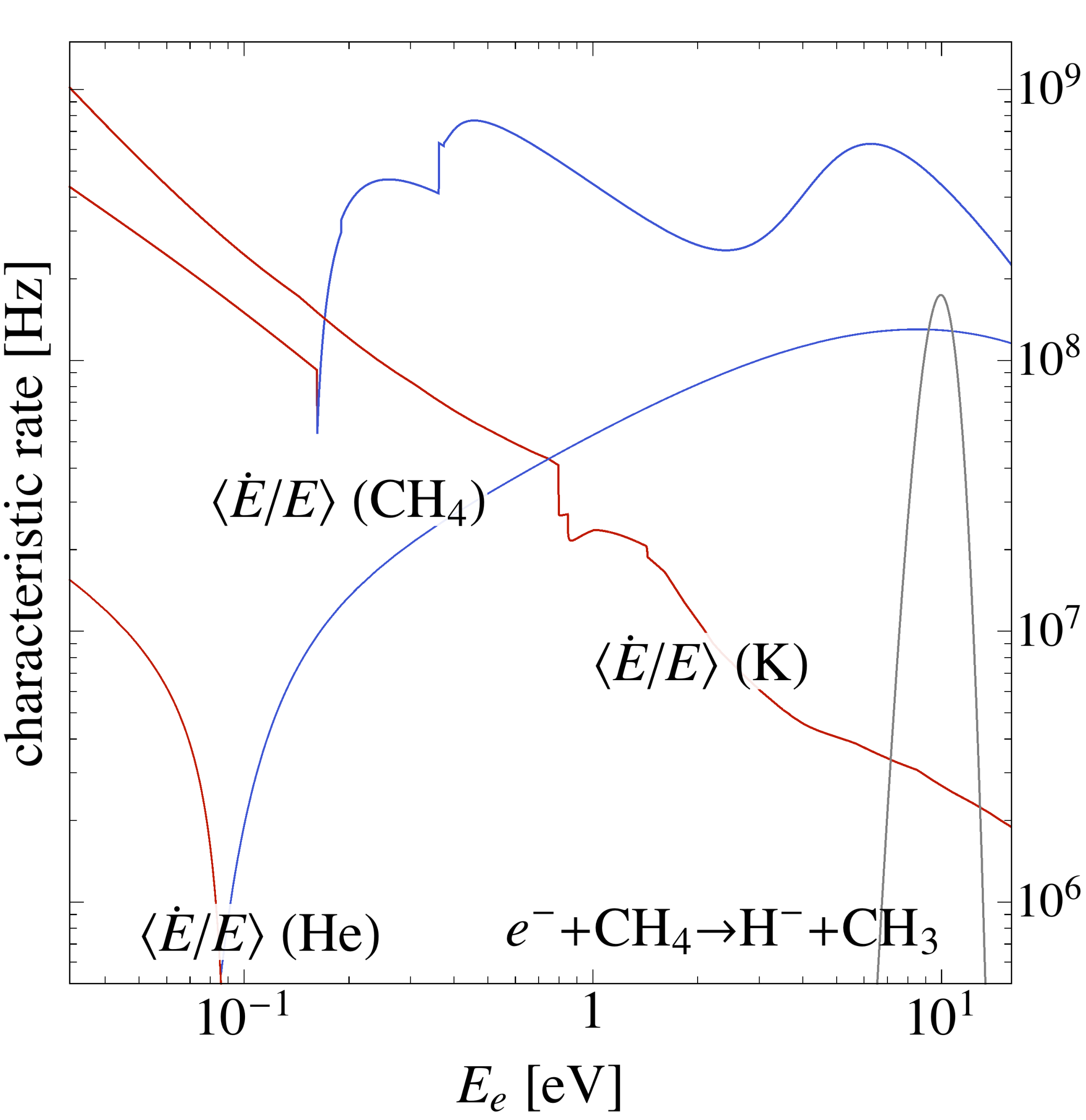}
      \caption{\label{fig:eTimes}
      For an electron with energy $E_e$, the main curves show the average gain (red), or loss (blue) rates -- as defined by $(dE/dt)/E$ -- due to each component in a ``typical'', active DPAL.  Note that the net effect of pumped potassium is purely energy gain.  The gray curve shows the rate for electrons to breakdown methane.}
   \end{figure}

   \subsection{Elastic and inelastic energy transfer (free electrons)}

   A new host of processes enter the picture once ions appear.  Fig. \ref{fig:eTimes} sketches gain and loss timescales for electrons in a typical potassium DPAL.  For the usual extent of population inversion, electrons experience net {energy} gain from collisions with potassium.  Inversion between D1 and D2 even gives a noticeable contribution, a fact previously recognized by Markosyan and Kushner\cite{2016_MarkosyanKushner_16CsDpalPlasma}, and Zatsarinny et al.\cite{2014_ZatsarinnyEtc_CsSigmas}.  Even if all the alkali were ionized, energy transfer via elastic collisions with helium still dominate { electron-electron energy transfer.  Energy transfer through inelastic} collisions with a hydrocarbon buffer can easily surpass {the energy transfer to} helium{} as fig.\ref{fig:eTimes} illustrates (cf. Section \ref{sec:A_EIT}).

   \subsection{Hydrocarbon breakdown mechanisms}

   Hydrocarbons may also present {a target for electron impact breakdown, or excited-alkali reactions}.  For methane specifically, ${\rm CH}_4 + {\rm e}^- \rightarrow {\rm CH}_3 + {\rm H}^-$ dominates electron-based breakdown\cite{2015_SongEtal_eIX_Methane}.  Data on subsequent pathways is sparse for heavy alkali, but Penning detachment predictions for other excited alkali-${\rm H}^-$ collisions\cite{1997_MartinBerry_eDetach_HnegxLiOrHe,1997_MartinBerry_eDetach_HnegxLiOrCa}, and measurements for oxygen anions striking excited oxygen molecules\cite{2008_MideyDV_OnegExcO2_BigPenDet} suggest that ${\rm KH}+{\rm e}^-$ formation using the abundant ${\rm K}^*$ will have cross sections of order $10^{-14}$cm$^2$.  Penning detachment using the ground state is much weaker, and has a high threshold\cite{2016_WuLYWJanev_HnegLi_LowPenDet}.

   Regarding { excited-state chemical} reactions, Azyazov et al.\cite{2017_AzyazovEt_ExcRbXM_H2orCH4orC2H6} excited Rubidium to its second resonance level, and attempted to measure both reactive, and non-reactive quenching with various molecules.  For methane in particular, they obtained a reactive branching ratio of $0.04\pm0.03$, and detected no RbH -- the most likely product expected based on their own theoretical calculations -- thus concluding { the reaction rate was negligible}.  { Given this limited, negative data for reactive pathways for hydrocarbon breakdown, we dot not include such processes for now.}

   \subsection{Recombination, and related dimer association rates}

   Regarding recombination, the introduction references have already noted a hierarchy where dissociative recombination (DR) dominates neutral- and electron- mediated three-body recombination (NMR, EMR), and radiative recombination trails far behind.  The more energetic, non-MB nature of the electrons renders recombination {-- especially three-body channels --} more difficult than previously expected.  Unfortunately, the dominant DR channel is still riddled with uncertainties in the energy-dependent cross section, product states, and auxiliary processes affecting the dimer ion population like association.  

   {As far as product states for DR, the rules of thumb relating potential crossings to favored product states \cite{1990_Mitchell_DRetcReview}, and analogies to experiments that do resolve products (e.g. Le Padellec et al.\cite{1999_LePadellecEt_DR1} for CN$^+$) indicate that of DR will likely be: one ground state atom, and one excited to the pooling levels, or levels just above.}

   Arimondo et al.\cite{1985_ArimondoEtc_DatapointCsDR} examined DR for argon-buffered cesium vapor at relevant temperatures and densities, and also reported a best fit rate for association, $\mathrm{Cs}^+ + \mathrm{Cs} + \mathrm{Ar} \rightarrow \mathrm{Cs}_2^+ + \mathrm{Ar}$, of $2.3\times10^{-23}$cm$^6$/s, but noted the error might be an order of magnitude.  To our knowledge, the nearest similar measurements are for $\mathrm{Cs}^+ + \mathrm{Cs} + \mathrm{Cs} \rightarrow \mathrm{Cs}_2^+ + \mathrm{Cs}$ at similar temperature, but lower buffer pressure, by Bergman and Chanin\cite{1971_BergmanChanin_CsXRates}, and Morgulis and Korchevoi\cite{1968_MorgulisKorchevoi}, who reported rates on the order of $10^{-30}$cm$^3$/s, and $10^{-26}$cm$^3$/s, respectively.


   The critical, and uncertain nature of the association rate warrants some brief technical discussion, especially to motivate the lower values that we later find necessary.  In a two-step termolecular reaction model\cite{2015_AttachmentRxnsBook}, the {\rm K} and {\rm K}$^{+}$ form a metastable $({\rm K}_2^{+})^*$ with some rate constant, $k_f$.  {This metastable state} dissociates in a characteristic lifetime, $\tau$, in the absence of stabilizing third-body collisions{, which have} characteristic rate constant $k_s = n_B\lrangSUscr{\sigma v}{\mboxss{stab}}{}$ where $n_B$ is the buffer density.  For an intermediate population in equilibrium, the effective association rate constant is:
   \begin{equation}
      k_{\mboxss{eff}} = \dfrac{ k_f k_s }{ 1/\tau + k_s }.
   \end{equation}
   The association rate reduces to $k_{\mboxss{eff}} \approx k_f$ when stabilization proceeds much faster than metastable disintegration, or $k_{\mboxss{eff}} \approx k_f k_s \tau $ in the opposite limit.  For an ion-induced dipole interaction\footnote{high ionization conceivably screens this interaction, but our estimates for the Debye length even when every alkali is ionized are on the order of 1000\AA}, the appropriate Langevin cross section just depends on (dipole) polarizability, $\alpha$, and collision energy, $E$, as $\sigma_L = \pi(2\alpha/E)^{1/2}$ (in atomic units).  Taking polarizabilities from Mitroy et al. \cite{2010_MitroyEt_Alphas}, leads to $k_f \approx 6\times 10^{-9}\mbox{cm$^3$/s}$.  Depending on whether the stabilizing atom ``sees'' the ion, and has a Langevin cross section to stabilize, or just a typical momentum transfer cross section, the stabilization rate will be $k_s \approx 10^{-12}\mbox{cm$^3$/s}$ or $k_s \approx 10^{-8}\mbox{cm$^3$/s}$.

   In order for the three-body rate constant in Arimondo et al. to be as large as $10^{-23}\mbox{cm$^3$/s}$, yet have the process scale with buffer density, the characteristic $\tau$ must be $10^{-3}$s or $10^{-7}$s depending on which limit of $k_s$ applies, but each value is roughly a thousand times greater than the respective $1/k_s$, contradicting the condition $\tau \ll 1/k_s$ to have buffer density dependence.  For an atmosphere or so of buffer, the $k_f \approx 6\times 10^{-9}\mbox{cm$^3$/s}$ estimated above coincides instead with a ``three-body'' rate around $\approx10^{-26}\mbox{cm$^3$/s}$.


   \subsection{Miscellaneous processes}

   Dissociative collisions by excited alkali, and electron impacts -- called dissociative excitation (DE) -- can counteract association.  We model the former process based on measurements for sodium by Tapalian and Smith\cite{1994_TapalianSmith_Na2p_dissoc} (Section \ref{sec:A_FIRST}).  Peak cross sections, and energy dependence for DE are fairly uniform among a variety of cation dimers\cite{2015_MotaponEt_DR1, 2011_LecointreEt_DR1, 2004_ElGhazalyEt_DR1, 1999_LePadellecEt_DR1}), so we also included a toy model of DE (Section \ref{sec:A_DRDE}), but found that it played a weak role.

   Neutral dimer association can divert population from the lasing cycle, {counteracted} by dissociation due to excited alkali\cite{2005_BanAP_Rb2_destr} (Section \ref{sec:A_FIRST}).  The neutral and ion dimer populations can interact directly through charge exchange, but at the moment we do not model this.

\section{Implementing the model}
\label{sec:NumIntro}

   Now we must simulate all the physics discussed above.  Processes involving just the thermalized, massive particles follow standard rate equations.  {For the basic three-level cycle, laser intensity, and pump term, we adopt the model of Hager and Perram\cite{2010_HagerPerram_3Lvl1}, which we summarize below.  To represent the non-MB electrons, we bin the spectrum to obtain ordinary differential equations, but this introduces issues with energy conservation which we {also address here, following which we explain how to resolve these numerical issues}.

   {We} model the core laser kinetics by source contributions ($\Delta s_x$) to densities of the first three levels, $n_i$ (starting at $i=1$ for the ground state), and the two-way laser photon density, $\psi_L$:
   \begin{align}
      \Delta s_1 &=  \sigma_{31}(n_3-2n_1)\omega + \sigma_{21}(n_2-n_1)\psi_L + n_2 \Gamma_{21} +n_3 \Gamma_{31}, \\
      \Delta s_2 &= -\sigma_{21}(n_2-n_1)\psi_L - n_2\Gamma_{21} +\gamma_{\mboxss{mix}}(n_3-2\exp[-\theta]n_2), \\
      \Delta s_3 &= -\sigma_{31}(n_3-2n_1)\omega - n_3 \Gamma_{31} -\gamma_{\mboxss{mix}}(n_3-2\exp[-\theta]n_2), \\
      \Delta s_L &= (rt^4\exp[2\sigma_{21}(n_2-n_1)l_g]-1)\psi_L/\tau_{\mboxss{RT}} + \lbrace \text{ $f\times n_1 A_{10}/\tau_{1}$ }\rbrace
   \end{align}
   where
   \begin{equation*}
      \omega = \dfrac{I_{\mboxss{P,in}}/h\nu_P}{\sigma_{31}(n_3-2n_1)l_g} \left(\exp[\sigma_{31}(n_3-2n_1)l_g]-1\right) t_P \left(1+t_P^2 r_P \exp[\sigma_{31}(n_3-2n_1)l_g]\right)
   \end{equation*}
   is the absorption rate based on input pump intensity $I_{\mboxss{P,in}}$.  The $g_i$ are level degeneracies, $A_{ij}$; radiative transition rates, $\Gamma_{ij}$; total decay rates, $\gamma_{\mboxss{fs}}$; the fine-structure mixing rate, $\tau_{\mboxss{RT}}$; the cavity round-trip time, $l_g$; the gain length, $r_x$ the output coupler reflection coefficients, $t_x$; the intracavity transmission coefficients (set to 1 throughout), $\theta=(E_3-E_2)/(\kboltz{}T)$, where $T$ and $\kboltz{}$ are temperature and the Boltzmann constant.  The last term in the laser equation is a rough estimate of the spontaneous emission seed for lasing, which we retain since our method is still based on evolving the equations in time, whether we apply to finding steady-state solutions or not.  The laser term contains a potentially very large exponential factor times the inverse cavity timescale.  We can address this extreme stiffness by keeping the overall effective timescale above $0.01\tau_{\mboxss{RT}}$ via
   \begin{equation}
      \dfrac{d\psi_L}{dt} \rightarrow a^{-1} \tanh\left[a\left(r_L \exp[2 \sigma \Delta{n} l_g] - 1 \right)\right]
   \end{equation}
   with $a=0.01$, for example.  Testing with different $a$, and the raw equations showed that this still preserved behavior on timescales of interest while avoiding situations where the code had to switch to excessively small timesteps.  We also track contributions to the general buffer specific energy density (the thermal bath) due to mixing of the D1 and D2 energy levels ($E_2$, $E_3$):
   \begin{equation}
      \Delta s_{\mboxss{eb}} = \gamma_{\mboxss{mix}} (n_3-2 n_2\,\exp[-\theta]) (E_3-E_2)
   \end{equation}

   {Note that energy exchange with the thermal bath also gets tallied for the other processes (like pooling, and quenching).}

   {Ensuring global energy conservation with a binned electron spectrum, and irregular grid, is not trivial.  Solutions} include: modifying the target-bin branching ratios to conserve energy at the cost of straying from the discretized differential cross section, and adopting a better, higher-order ``sub-grid'' model than a flat-top distribution {(see e.g. Le and Cambier\cite{2017_HaiJl_SubGrid})}.  The sub-grid approach avoids unsavory tweaks to the cross sections, and better handles transitions with energy less than a bin width.  For now though, re-weighting is {simpler} to implement, and {test various methods with}, but {the higher order discretization warrants further} consideration.  {This accounts for ionization, recombination, and inelastic collisions for free electrons,} which just leaves energy transfer with the buffer.  We base our formulation on the Lorentz model with a correction for the finite temperature of the buffer particles, (see e.g. Loureiro \& Amorim \cite{2016_LoureiroAmorimBook} where it is given in terms of velocity magnitude)
   \begin{equation}
   \label{eqn:drag1}
      \partial_t n_E = \partial_{E} \left[ \dfrac{E-\kboltz{}T/2}{\tau_{\mboxss{exch}}(E)} n_E + \dfrac{E\,\kboltz{}T}{\tau_{\mboxss{exch}}(E)} \partial_{E} n_E \right],
   \end{equation}
   where $E$ is electron energy, $n_E$ the electron differential energy density, and $\tau_{\mboxss{exch}}$ the characteristic energy exchange timescale.  We implement the discretized version as a series of upwinded advection fluxes between neighboring bins, plus a diffusion term, so the resulting source terms are more automatically conservative.  See Section \ref{sec:A_drag} for details.  

   \subsection{Conservation-enforcing projection ala Sandu (2001)}
   \label{sec:SanduProj}

   Resolving trace populations while avoiding negative densities and conservation error is {a familiar} challenge, motivating various approaches over the years.  {For example,} Preussner and Brand\cite{1981_PreussnerBrand}, and Bertolazzi\cite{1996_Bertolazzi_PosConsKins} {both} focus on preserving non-negativity with semi-implicit and implicit methods respectively.  Instead of focusing on the integration method, Sandu\cite{2001_Sandu_PosProj} remarks that finding a point obeying conservation rules, closest to some arbitrary method's guess, just defines a linearly-constrained quadratic optimization problem.  This {approach is highly general}, avoids longterm drift in error, and {forms} the backbone of our code.

   The projection relies on expressing conservation laws as a linear constraint:
   \begin{equation}
      A^{\intercal} {\bf n}' - A^{\intercal} {\bf n}_0 -{\bf x} = 0
   \end{equation}
   where ${\bf n}_0$ is some previous, trusted population vector, and ${\bf x}=0$ absent any external source/sinks.  Sandu reviews how to find $A$ given any stoichiometry, but alkali number and charge conservation rows can be written by inspection for our problem:
   \begin{alignat}{9}
      A_N^{\intercal} &= (&1,\hdots,1,&\quad && 2,\quad && 1,\quad && 2,\quad &&~1,~1,\hdots\quad &0,\hdots&), &&\mbox{ and }A_N^{\intercal} {\bf n} = n_{\mboxss{alk}}(t_0) \\
      A_Q^{\intercal} &= (&0,\hdots,0,&\quad && 0,\quad && 1,\quad && 1,\quad &&-1,-1,\hdots\quad &0,\hdots&), &&\mbox{ and }A_Q^{\intercal} {\bf n} = 0 \\
          & &\mbox{\small ${\rm K}({\rm nLJ})$}& ~ &&\mbox{\small ${\rm K}_2$} ~ &&\mbox{\small ${\rm K}^{+}$} ~ &&\mbox{\small ${\rm K}_2^{+}$} ~ &&\mbox{~e$^-$ {\small bins}} ~ &\mbox{\small misc.}& &  \nonumber
   \end{alignat}

   {In principle, neutral buffers will contribute their own rows, and internal energy conservation can be included in a similar fashion.  In practice, buffer conservation holds well either way, while adding internal energy means adding a large external source term (the pump).}  For large timesteps, the energy conservation correction tends to interfere with the others, so some numerical error from the (time) integration method can still affect energy for now.  This is not an issue for steady-state-seeking simulations, as any equilibrium is conservative by construction of the source terms.

   This $A$ then features in the optimization problem
   \begin{equation}
      {\bf n}^{n+1} = \mathrm{argmin} \left(\dfrac{1}{2} {\bf n}^{n+1} (G {\bf n}^{n+1}) - (G \tilde{\bf n}^{n+1})^{\intercal} {\bf n}^{n+1} \right) : A^{\intercal}{\bf n}^{n+1}=y_0, {\bf n}^{n+1}\geq \eps.
      \label{eq:SanduOptProblem_WithG}
   \end{equation}
   where $\tilde{\bf n}^{n+1}$ is the solver's uncorrected guess, and $G$ is the error metric for the numerical solver.  In Sandu's example $G$ is a diagonal matrix
   \begin{equation*}
      G_{ii} = 1/\left(N_{\mboxss{species}}(\mathrm{tol}_{\mboxss{abs}}+\mathrm{tol}_{\mboxss{rel}}|n_i|)^2\right).
   \end{equation*}
   The weighting assumes extra importance in a problem like ours with many levels of the same atom (i.e. a long row of zeros in $A^{\intercal}$).  Without any for example ($G_{ij}\equiv\delta_{ij}$), the sea of trace populations will ebb and flow when the projection corrects changes in large, dynamic populations.

   This tool frees us to take steps $\Delta{\bf n}$ using a simple integration method like linearized implicit Euler:
   \begin{equation}
      (I-\Delta{t} J)\Delta{\bf n} = {\bf s} \Delta{t}
   \end{equation}
   where ${\bf s}$ is the total source vector, and $J$ the Jacobian.

   \subsection{Alternative approaches}
   \label{sec:AltMethods}

   We consider it worth {mentioning here some} alternative approaches, and immediate issues we faced applying them.

   Computational singular perturbation (CSP)\cite{1994_LamGoussis_CSP, 2005_ValoraniEt_CSP} takes the opposite approach: instead of addressing accuracy issues for implicit methods, it addresses stiffness issues for explicit ones.  It achieves this by breaking up the source term into modes with characteristic timescales, typically found by straight-forward eigen-decomposition, or block diagonalization of the Jacobian, with a possible higher order correction for evolution of the bases.  `Fast' modes can reach a quasi-equilibrium with respect to the slow ones (or `exhaust').  For example, the laser levels in our problem often reach a quasi-equilibrium that evolves slowly with respect to the slow changes caused by pooling, etc.  CSP works well on the `conventional' and three-level kinetics of our problem, but transfer between the electron bins alters modes on a fast (Courant-Friedrichs-Lewy) timescale, generating fast modes which do not exhaust for a long time without the (expensive) higher order corrections.  Simple attempts at splitting off processes that only shift electrons in energy helped little as the CSP modes still changed drastically between steps.  Efficient application of CSP would likely require representing the electron spectrum differently, e.g. via moments, or some other coarser expansion.

   Applying the semi-implicit algorithm in Preussner and Brand is straightforward, and should complement the projection step nicely: it avoids negative densities that increase the cost of the projection, while the projection should reduce the need for very small steps to ensure conservation over long times.  However, this approach still required small timesteps to avoid oscillations associated with the radiative component, and to a lesser extent, electron drag and excitation/de-excitation.  Given the lower cost per step, some variation that splits off the radiative component may still offer a practical path forward for larger, more expensive engineering simulations

\section{Applying the model}
\label{sec:Applying}

   Knowing the physical ingredients and main uncertainties, numerical techniques and issues, we can now compare predictions to experiments in a time-dependent, ionization-prone regime, as well as draw some general lessons from simulating general, steady-state situations.

   \subsection{Testing against time-dependent experiments}
   \label{sec:Applying_ZRSK17}

   To avoid complications from thermal build-up and other slow processes, Zhdanov et al. \cite{2017_ZhdanovEt_KDPAL_BufferStudy} ran a set of experiments with time-varying, sub-millisecond pump pulses for a potassium vapor at 190$^{\circ}$C with varying buffer composition.  {Their main} experiments used 500 Torr of buffer with a varying percentage of methane, and a peak pump power of 160W, translating to an intensity around 30kW/cm$^2$ for their reported beam profile.  They reported laser output, as well as 5\,$^2$P$_J$-4\,$^2$S$_J$ fluorescence, but not in absolute units.  They also measured 5\,$^2$P$_J$ fluorescence for trials with 200 Torr of pure helium, and pure argon, and saw that neither composition lased.

   We first ran the model with just the basic three-level model turned on to make sure it reproduced the lasing threshold, using the reported pump linewidth, laser beam profile, gain and cavity lengths, and pressure-broadening of the D1, and D2 transitions based on Pitz et al.\cite{2014_PitzEt_K4P_Broadening} (c.f. Section \ref{sec:A_FIRST}).     
   Initial runs with the default kinetics, and slight variations, established some basic tenets of an `alkali-depletion' paradigm under these conditions:
   
   \begin{itemize}
      \item[ 1 ] \label{ten:EIIvsDR} past $\sim$1-10\% (alkali) ionization, impact ionization and DR dominate free electron {population gain and loss}
      \item[ 2 ] \label{ten:Sigmoid} this leads to sigmoidal growth of ion fraction, where drag, and pump intensity can also skew its shape, saturation timescale, and final ion fraction
      \item[ 3 ] \label{ten:SatMech} at late times, the barrier to more ionization is re-energizing low energy, post-impact electrons back above ionization thresholds before {they recombine}
      \item[ 4 ] \label{ten:5Ptrend} increased buffer drag delays and diminishes the peak in 5\,$^2$P$_J$ population matching the trend of their 200 Torr experiments, but the early growth is never so linear in time as seen in the experiments
   \end{itemize}

   \begin{figure}[!htbp]
      \includegraphics[width=0.99\textwidth]{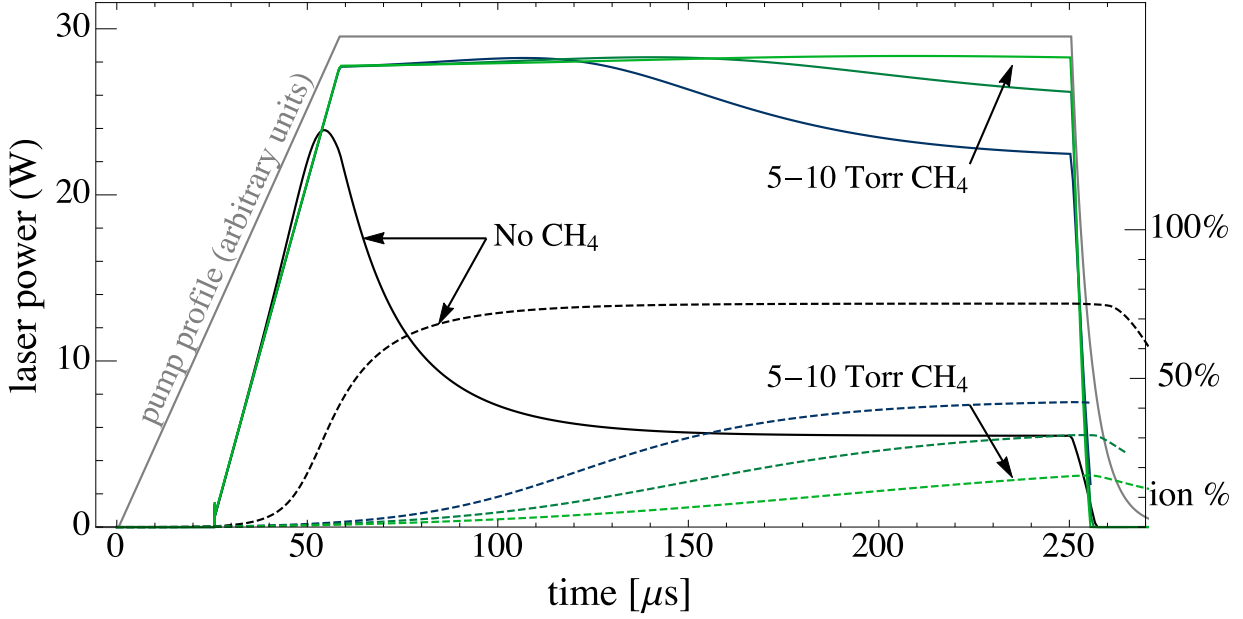}
      \caption{\label{fig:RefRuns1} Solid curves show laser power output for pure helium and various amounts of methane where {\it total} pressure remains 500 Torr for all cases.  Dashed curves show ionization fraction with scale on the right axis, and the gray curve shows the pump profile in arbitrary units.}
   \end{figure}

   The simulations in fig. \ref{fig:RefRuns1} illustrate the basic depletion mechanism, where the association rate for K$_2^+$ production was set to the equivalent of a $10^{-26}$cm$^6$/s three body rate, and atomic-level quenching was reduced to 20\% of its default value.  Since we assumed it most significantly quenches the 5\,$^2$P$_J$, 6\,$^2$S$_J$, and 4\,$^2$D$_J$ levels, further reduction of the DR cross section, or faster impact ionization out of these levels would generate similar behavior.

   \begin{figure}[!htbp]
      \includegraphics[width=0.99\textwidth]{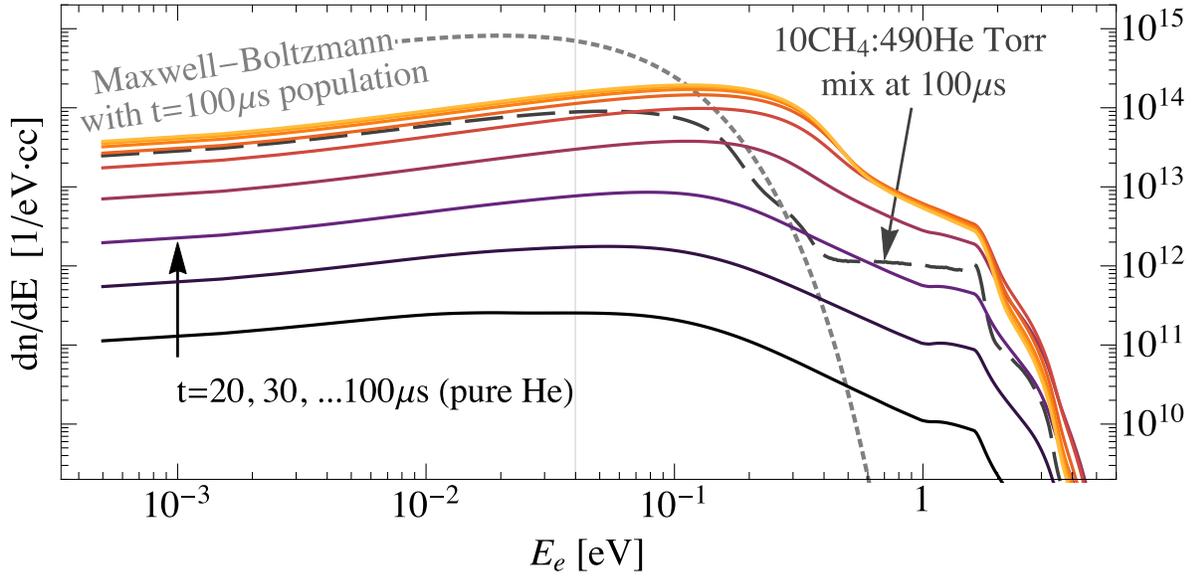}
      \caption{\label{fig:espec_example} Dark to light solid curves show 10$\mu$s steps in electron spectrum growth for the pure helium simulation of fig.\ref{fig:RefRuns1}.  The dashed curve shows the spectrum for the 10:490 methane:helium mixture at late times.  For contrast, the dotted curve shows a Maxwell-Boltzmann distribution with the gas temperature, and same total density as the last pure-helium curve.}
   \end{figure}

   Fig. \ref{fig:espec_example} shows the evolution of the electron energy distribution in the pure helium run, a spectrum for one of the methane-added runs, and a buffer-temperature Maxwell-Boltzmann distribution with the same final electron count as the pure helium case.  The non-MB curve is roughly one-tenth the MB curve below 0.1eV, so already the total recombination rate is significantly changed.  The curve for the methane case highlights methane's ability to suppress not only number, but mean electron energy.  Figures \ref{fig:4SProcMags}-\ref{fig:D2ProcMags} break down various contributions to the source terms for the ground state, D1, and D2, which demonstrates the role of impact ionization here, and explains a strange phenomenon described below.

   \begin{figure}[!htbp]
      \begin{subfigure}[t]{0.449\textwidth}
         \includegraphics[width=1.00\textwidth]{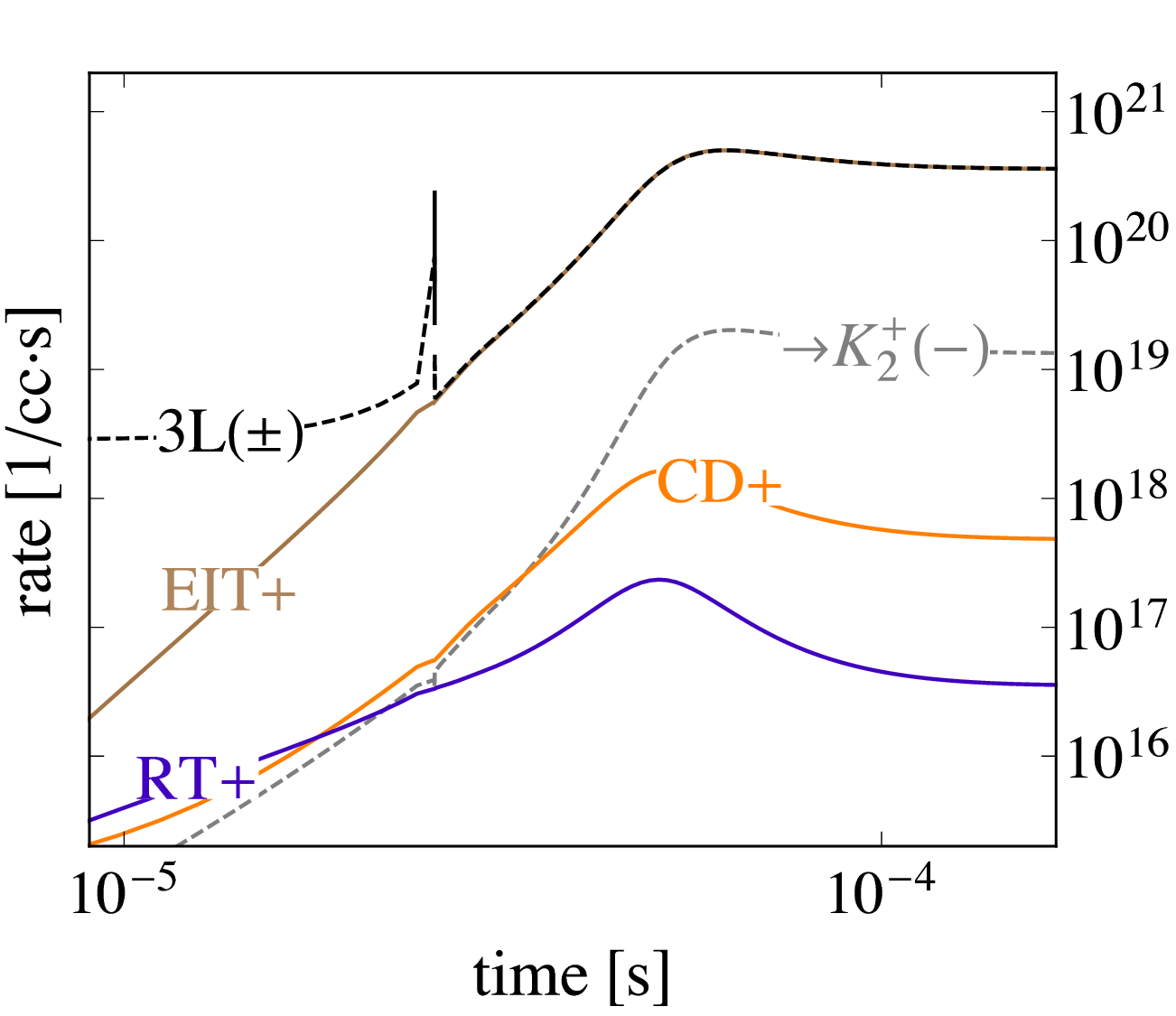}
      \end{subfigure}
      \begin{subfigure}[t]{0.449\textwidth}
         \includegraphics[width=1.00\textwidth]{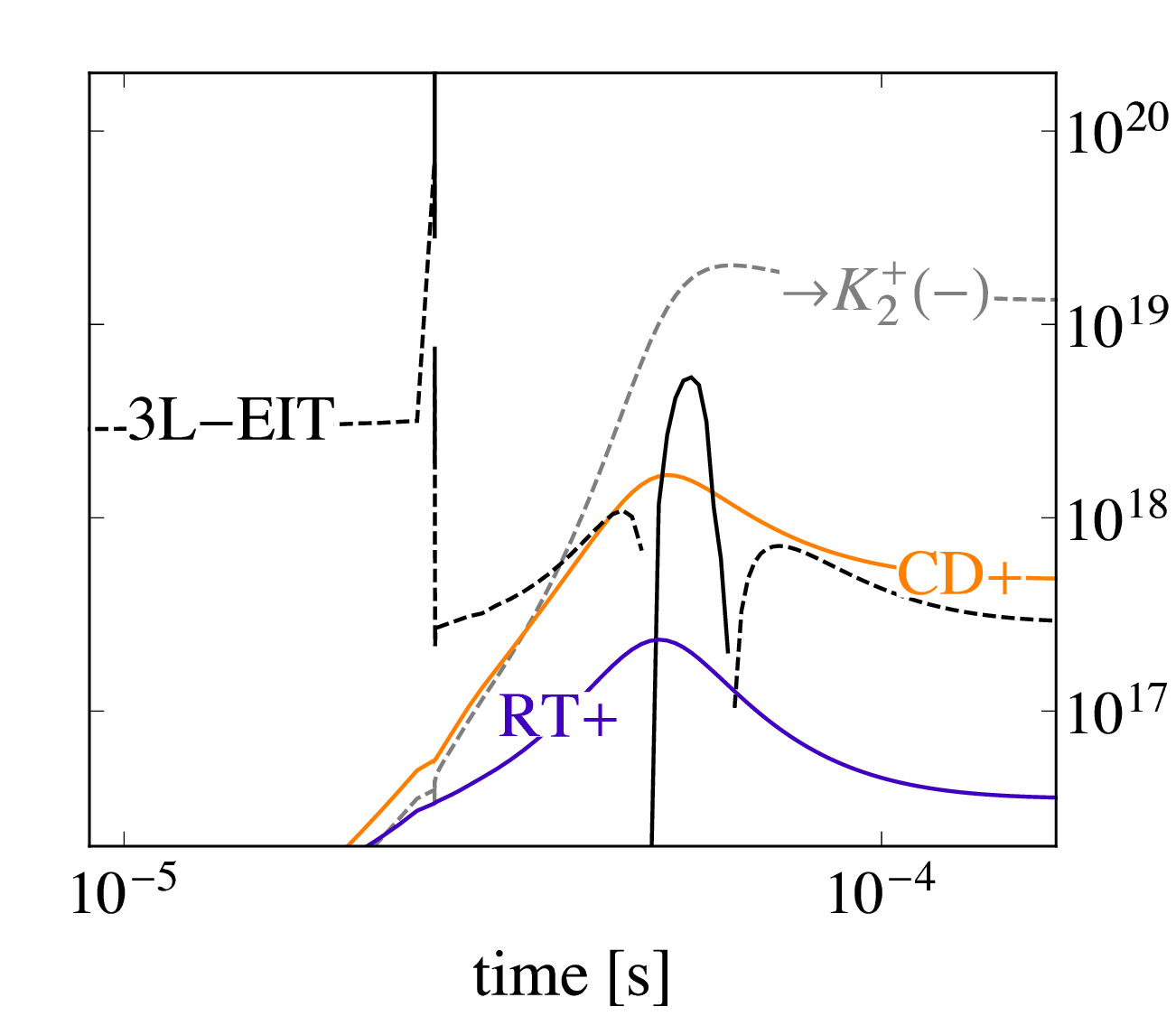}
      \end{subfigure}
      \caption{\label{fig:4SProcMags} Contributions/subtractions (solid/dashed) to the ground state from the sum of the basic three-level source terms (3L), EIT, radiative transfer (RT), collisional dissociation (CD), and association with K$^+$ forming K$_2^+$.  The right panel adds EIT to three-level losses to highlight the rate of net loss via association}
   \end{figure}
   \begin{figure}[!htbp]
      \begin{subfigure}[t]{0.449\textwidth}
         \includegraphics[width=1.00\textwidth]{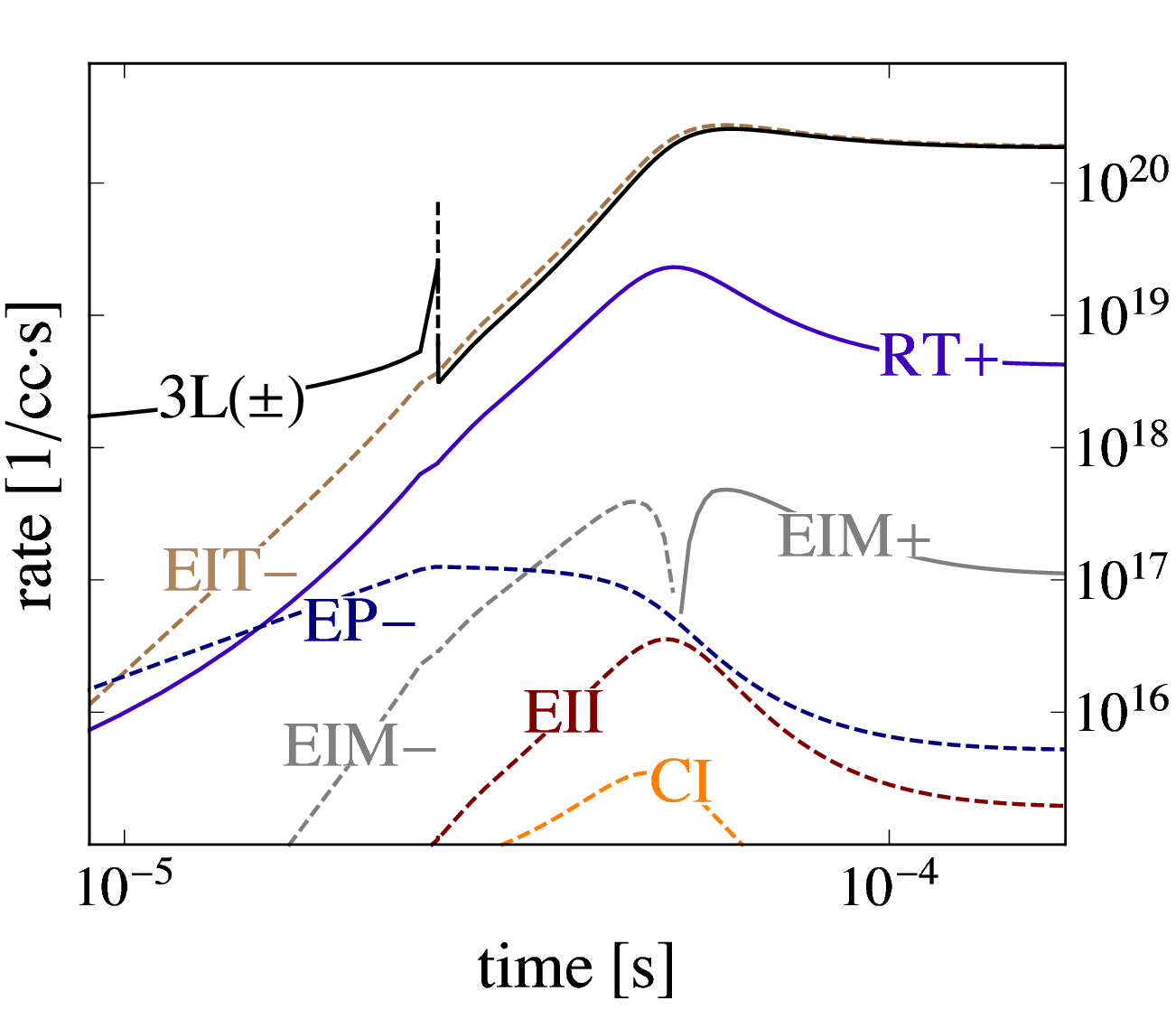}
      \end{subfigure}
      \begin{subfigure}[t]{0.449\textwidth}
         \includegraphics[width=1.00\textwidth]{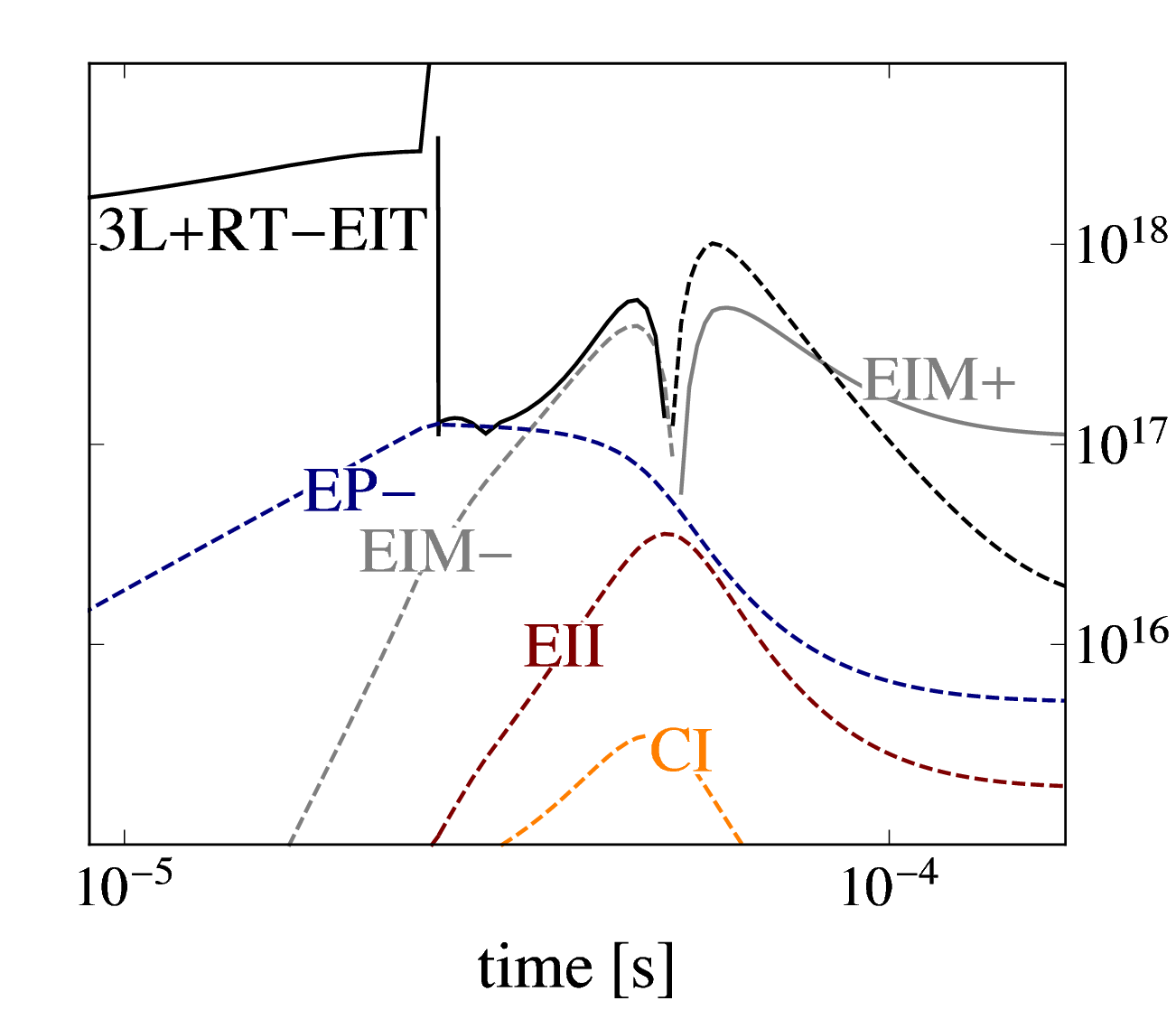}
      \end{subfigure}
      \caption{\label{fig:D1ProcMags} Same as fig.\ref{fig:4SProcMags}, but for D1, so energy pooling (EP), collisional ionization (CI), impact ionization (EII), and impact mixing (EIM) are relevant.  This time radiative decays from levels above are included in the total for major processes.  For the simulation parameters, impact fine-structure mixing has less affect on long term change in gain than the association rate in fig.\ref{fig:4SProcMags}.}
   \end{figure}
   \begin{figure}[!htbp]
      \begin{subfigure}[t]{0.449\textwidth}
         \includegraphics[width=1.00\textwidth]{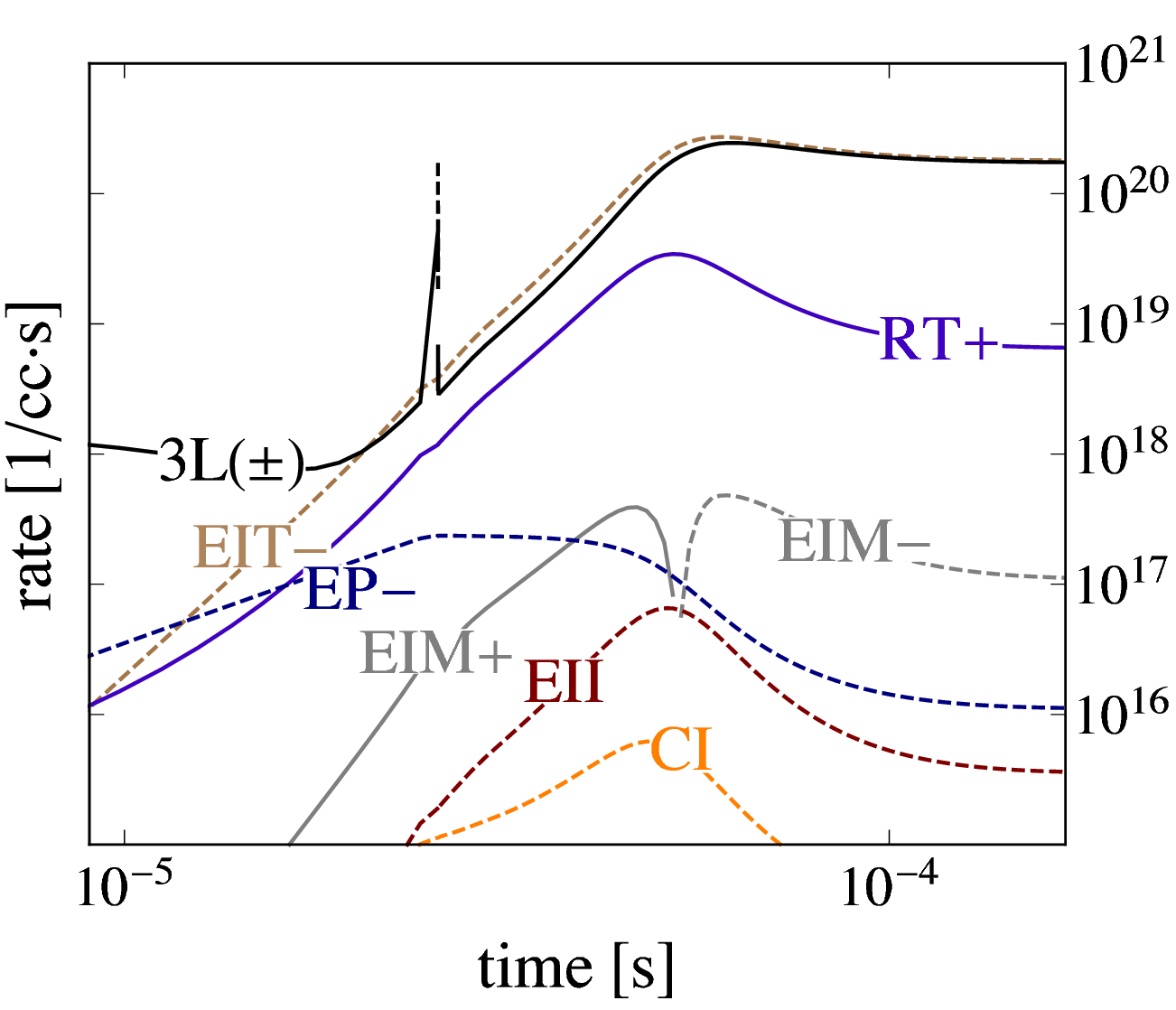}
      \end{subfigure}
      \begin{subfigure}[t]{0.449\textwidth}
         \includegraphics[width=1.00\textwidth]{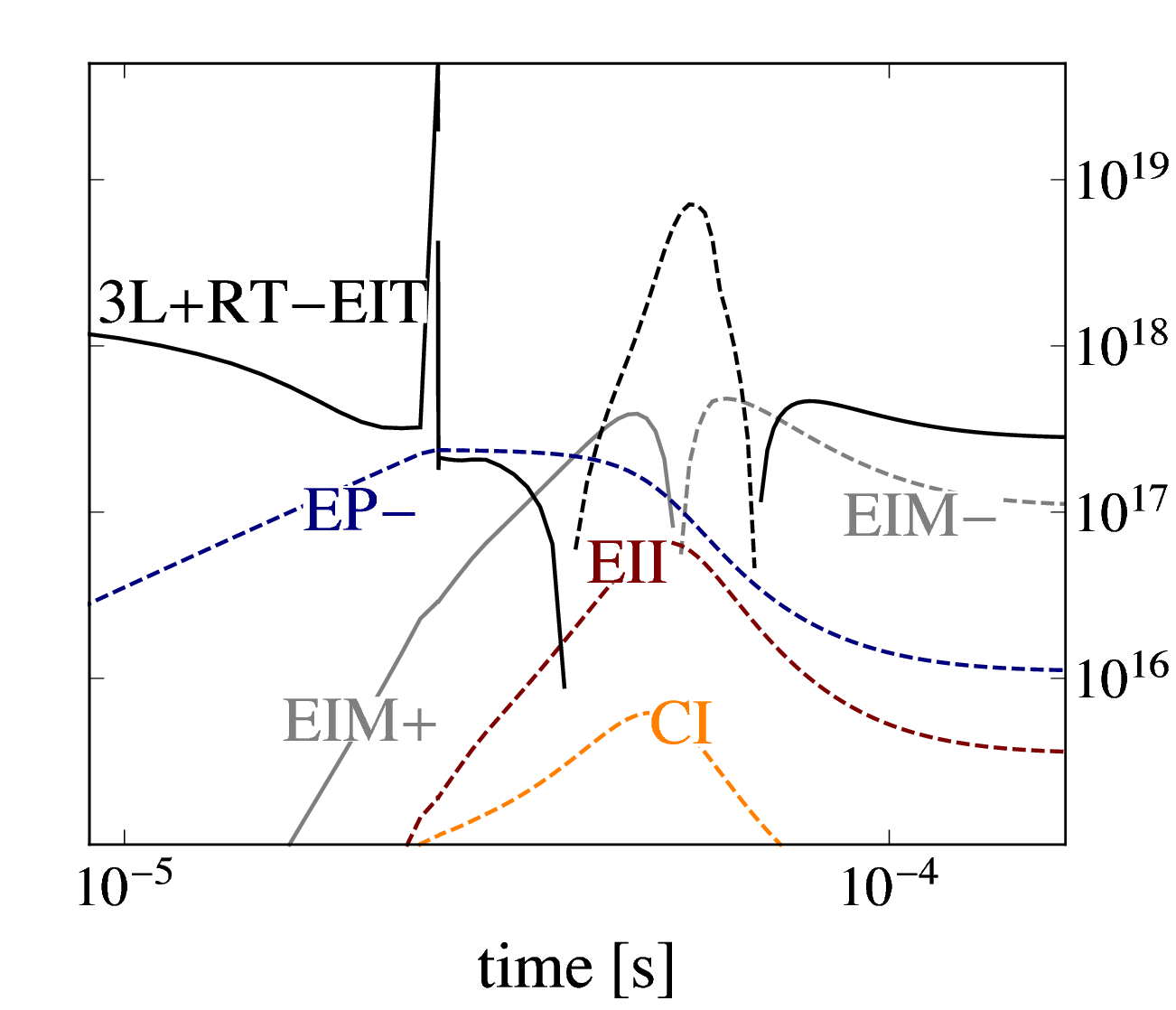}
      \end{subfigure}
      \caption{\label{fig:D2ProcMags} Same as fig.\ref{fig:D1ProcMags}, but for D2.}
   \end{figure}

   For pure helium, the rise in laser output slightly overshoots the other mixtures before cresting in fig.\ref{fig:RefRuns1}, and the source term figures reveal how moderate ionization briefly boosts laser gain to cause this.  From lasing onset -- seen as a spike around 20-30$\mu$s -- to 50$\mu$s or so into the simulation, the sum of three-level, impact transfer processes, and eventually association into ${\rm K}_{2}^+$ all drive significant ground state depletion.  Meanwhile, D1 experiences net growth, thus raising laser gain until severe ionization depletes both populations.  Net electron impact mixing of the D1, and D2 levels also switches sign around this time, as D2 atoms sufficiently outnumber D1 atoms.

   Artificially raising {absorption} cross sections by a factor of 5 dramatizes this effect in fig.\ref{fig:ErrRuns1}.  This also leads to delayed, saturated curves for the methane mixtures resembling their output curves in the experiment, but if this mechanism does play a role in explaining the responses with methane, it can not be the sole cause.  For now, this overshoot serves as a hint for where we should examine the physics and methods further, e.g. perhaps the modeled electrons are too hot at early times leaving excitation too strong versus de-excitation, or perhaps the ${\rm K}_{2}^+$ association rate is still too high.

   \begin{figure}[!htbp]
      \includegraphics[width=0.99\textwidth]{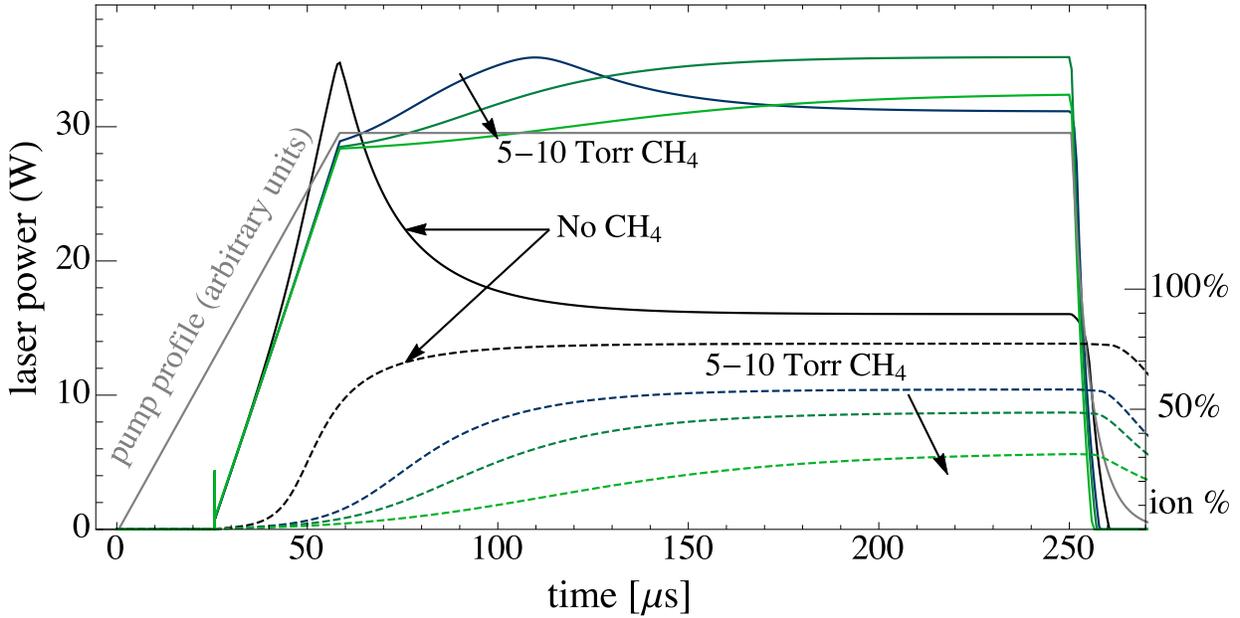}
      \caption{\label{fig:ErrRuns1} Same as fig.\ref{fig:RefRuns1}, but with artificially higher absorption cross sections.  Overshoot during methane runs leads to the type of delayed \& saturated curves seen in experiments, except the trend of final laser output falling with methane density starts too soon, and it overshots too much in the pure helium trial.}
   \end{figure}

   \subsection{Ionization can drive significant heat loading}
   \label{sec:Applying_Heat}

   Predictions of vigorous heat generation by free electrons in a pumped, alkali vapor date at least as far back as a paper by Measures\cite{1970_Measures_EarlyKplas}.  Additional, unexpected heating can hurt beam quality, among other effects, but this aspect of ionization receives relatively less attention, so we review some simple arguments behind strong heating.

   From fig. \ref{fig:eTimes}, one sees that a typical electron born around 0.3-0.5 eV, will frequently gain 1.6eV by de-exciting a K(4P) atom on a sub-\mbox{$\mu$s} timescale.  Often after one such jump, energy loss to the buffer is suddenly much faster.  As ionization and recombination are orders of magnitude slower, this means a typical electron can convert a pump photon worth of energy to the buffer hundreds of times while free.  

   Meanwhile, bound electrons converting the $\approx 0.01$eV D1-D2 gap to heat at a 1-10 GHz rate via fine-structure mixing present the greatest heat source in ionization-free models, and the logical point of comparison:
   \begin{equation}
      f_{\mboxss{ioni}} n_A \mbox{(1 eV)} \mbox{(1 MHz)} \mbox{   vs.   }
      n_A \mbox{(0.01 eV)} \mbox{(1 GHz)}.
   \end{equation}
   This shows the drag heat load approaches the inherent mixing one around 10\% ionization.

   Electron energy loss to methane will largely go into local heat as well: Menard-Bourcin et al. \cite{2005_MenardBourcinEt_EQVT_CH4xVar} give cross sections for vibrational-translational energy transfer for methane in helium, and the resulting rates exceed radiative decays based on Yurchenko et al. \cite{2013_YurchenkoEt_VibrFoscCH4} by several orders of magnitude.  The code accounts for this.

   Fig. \ref{fig:Starfish} shows heat load, and fraction of it from mixing, helium energy transfer, methane energy transfer, atomic level quenching, and other sources, for a small grid of alkali density, and methane buffer fraction at a fixed value of angle-averaged and frequency-integrated pump intensity, $\bar{J}$ of 100kW/cm$^2$.  At moderate alkali density, adding a small amount of methane drops ionization and associated heat load substantially.  For the same total buffer density at higher alkali densities, methane does little besides replace helium as the pathway for heat generation.  Total buffer density must be raised for such conditions.
   \begin{figure}[!htbp]
      \includegraphics[width=0.80\textwidth]{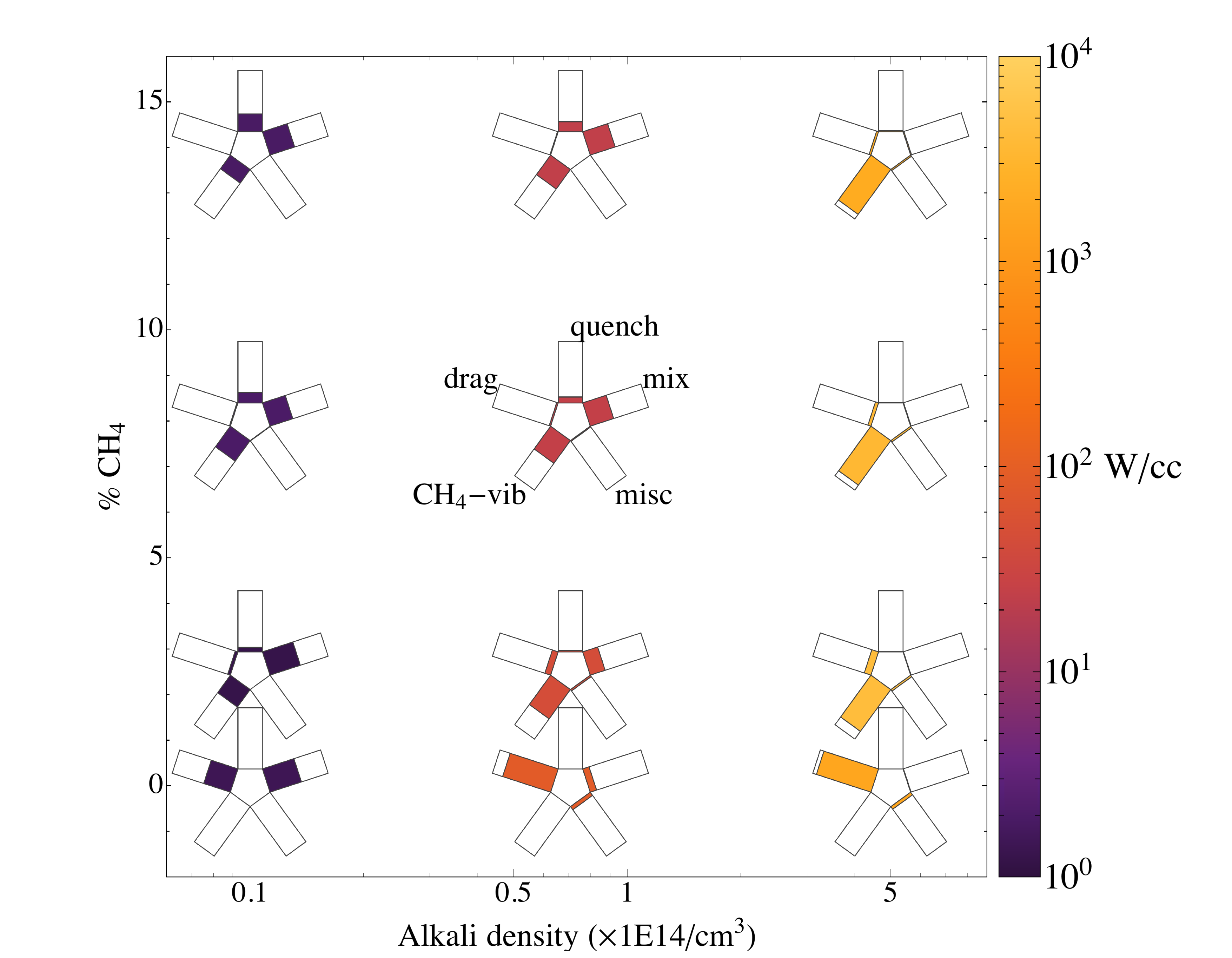}
      \caption{\label{fig:Starfish} ``Starfish'' arms show relative contributions from (4\,$^2$P$_J$) fine-structure mixing, potassium level quenching, energy transfer to helium, energy transfer to methane (when present), and miscellaneous processes according to the legend, while their hue indicates total heat load in Watts per cubic centimeter.  The buffer pressure was one atmosphere in all these simulations.}
   \end{figure}

\section{Conclusions}
\label{sec:Conc}

   Active DPAL conditions fundamentally lead to a non-Maxwell-Boltzmann distribution for free electrons, and ignoring this effect underestimates the propensity for alkali to ionize, as well as the extra heat load associated with ionization.  By comparing the model to experiments performed by Zhdanov et al., whose variations in time and composition probed the transition from good to poor performance, we confirmed this basic point, but also showed important discrepancies.  The model predicted a slight, temporary boost to laser gain at early times for the pure helium case, in contrast to the experiment.  The model also over-predicted the efficacy of methane.  The latter is less surprising in the face of quenching uncertainties, while the former constrains more central aspects of the model, like impact ionization, drag, alkali dimer ion association. 

   To reduce modeling uncertainties, we have started implementing the higher-order discretization of the distribution, obtaining more accurate impact ionization estimates, and tying the higher fidelity kinetics to a more detailed spatial model of the laser.  We also intend to investigate three-body recombination with non-MB electrons more rigorously.

   The comparison illustrates the utility of time-dependent pumping on kinetics-relevant timescales for setting stronger constraints on models.  Further variations on this idea, as well as more quantitative diagnostics for the basic three levels and excited states, are warranted.  Direct measurements, or higher-quality estimates of the most uncertain processes would obviously be most helpful too.


\begin{acknowledgments}

   We thank Dr's. Habib Najm, John Shadid, and their colleagues for helpful discussions, especially regarding CSP.  HJC thanks Ben Oliker for fielding basic questions at the start of the project.  We also thank an anonymous referee for detailed feedback, which significantly improved the paper.

\end{acknowledgments}

\appendix

   \section{Neutral and heavy ion kinetics}
   \label{sec:A_Heavies}
   \label{sec:A_FIRST}

   We included all individual fine-structure levels from 4\,$^2$S$_{1/2}$ through 6\,$^2$S$_{1/2}$, and used radiative transition rates from calculations by Nandy et al.\cite{2012_NandyEtal_K_tijEtc} except for D1 and D2 transitions to ground.  Processes involving free electrons do not have rates, but energy-dependent cross sections.  Details, or references appear in subsequent appendices where not already given.  

   For the basic three levels, the mixing rate $\gamma_{\mboxss{mix}}$ was based on
   \begin{equation}
      \gamma_{\mboxss{mix}} = (n_{\mboxss{He}} Q_{\mboxss{He}} + n_{\mboxss{CH$_4$}} Q_{\mboxss{CH$_4$}}) \left\langle v \right\rangle_{\mboxss{th}}
   \end{equation}
   with $Q_{\mboxss{He}} = 18.7\AA$, $Q_{\mboxss{CH$_4$}} = 58.9\AA$, and $\left\langle v \right\rangle_{\mboxss{th}}$ the mean thermal velocity.  Broadening rates for D1, and D2 absorption were taken from Pitz et al.\cite{2014_PitzEt_K4P_Broadening}:
   \begin{equation}
      \Gamma_{\mboxss{D1}}/\mbox{MHz} = 19.84\dfrac{ P_{\mboxss{He}} }{\mbox{Torr}} + 13.08\dfrac{ P_{\mboxss{CH$_4$}} }{\mbox{Torr}}\mbox{, and}
   \end{equation}
   \begin{equation}
      \Gamma_{\mboxss{D2}}/\mbox{MHz} = 19.84\dfrac{ P_{\mboxss{He}} }{\mbox{Torr}} + 27.78\dfrac{ P_{\mboxss{CH$_4$}} }{\mbox{Torr}}\mbox{.}
   \end{equation}
   We used these to calculate line-center cross sections, and overlap with the pump spectral profile, which Zhdanov et al. said was a Gaussian with 12\,GHz full-width half-maximum.

   As noted, our pooling rates our informed by the results of Namiotka et al.\cite{1997_NamiotkaHA_EP_K}.  We did not make any assumptions about hybrid-$J$ (e.g. 4\,$^2$P$_{3/2}$+4\,$^2$P$_{1/2}$) pooling rates, but know from other alkali\cite{1996_VadlaEtc_CsPooling} that they are likely comparable, so we expect the model to underestimate pooling.
   \begin{table}[!htbp]
      \caption{\label{tab:EPs} Energy-pooling rate constants in cm$^3$/s; $J$ denotes each fine-structure level}
      \begin{ruledtabular}
      \begin{tabular}{cccccccc}
      \multicolumn{1}{c}{reactants} & excited product & rate & \multicolumn{1}{c}{reactants} & excited product & rate \\
      \hline
      4\,$^2$P$_{3/2}$ , 4\,$^2$P$_{3/2}$ & 5\,$^2$P$_{J}$ & $4\times10^{-11}$ &
         4\,$^2$P$_{1/2}$ , 4\,$^2$P$_{1/2}$ & 5\,$^2$P$_{J}$ & $9.7\times10^{-11}$ \\
      4\,$^2$P$_{3/2}$ , 4\,$^2$P$_{3/2}$ & 6\,$^2$S$_{J}$ & $8.2\times10^{-12}$ &
         4\,$^2$P$_{1/2}$ , 4\,$^2$P$_{1/2}$ & 6\,$^2$S$_{J}$ & $2.7\times10^{-12}$ \\
      4\,$^2$P$_{3/2}$ , 4\,$^2$P$_{3/2}$ & 4\,$^2$D$_{J}$ & $2.0\times10^{-11}$ &
         4\,$^2$P$_{1/2}$ , 4\,$^2$P$_{1/2}$ & 4\,$^2$D$_{J}$ & $1.2\times10^{-11}$ \\
      \end{tabular}
      \end{ruledtabular}
   \end{table}

   The Earl and Herm\cite{1974J_EarlHerm_EQ_NaKxMs} quenching cross section for 5\,$^2$P$_{J}$ by methane was not fine-structure specific, nor was the destination state resolved.  For now, we just took a similar multiplet-to-multiplet cross section for the neighboring states with similar energy separation, and divided it by the number of levels involved, and scaled the rates up or down all together.  For the simulations shown, we used the cross sections in table \ref{tab:EQs}.

   \begin{table}[!htbp]
      \caption{\label{tab:EQs} Default, hard-sphere methane quenching cross sections; $J$ and $J'$ denote each possible fine-structure level}
      \begin{ruledtabular}
      \begin{tabular}{cccccc}
      initial & final & $\sigma (\AA^2)$ & initial & final & $\sigma (\AA^2)$ \\
      \hline
      5\,$^2$P$_{J}$ & 3\,$^2$D$_{J'}$ & $2$ & 
         5\,$^2$P$_{J}$ & 5\,$^2$S$_{J'}$ & $2$ \\ 
      4\,$^2$D$_{J}$ & 5\,$^2$P$_{J'}$ & $3$ & 
         6\,$^2$S$_{J}$ & 5\,$^2$P$_{J'}$ & $3$ \\ 
      4\,$^2$P$_{J}$ & 4\,$^2$S$_{J'}$ & $6$ & 
         & & \\
      \end{tabular}
      \end{ruledtabular}
   \end{table}

   As discussed above, the forward dimer ion association rate was still modeled as an effective three-body rate
   \begin{equation}
      {\rm K}^+ + 4\,^{2}{\rm S}_{1/2} + He \overset{k}{\rightarrow} {\rm K}_2^+ + He \mbox{; $k=10^{-26}$ cm$^6$/s}.
   \end{equation}
   We took the neutral association rate to be
   \begin{equation}
      2\times 4\,^{2}{\rm S}_{1/2} + He \overset{k}{\rightarrow} {\rm K}_2 + He \mbox{; $k=10^{-31}$ cm$^6$/s}.
   \end{equation}

   Collisional dissociation by excited atoms, again based on Tapalian and Smith\cite{1994_TapalianSmith_Na2p_dissoc}, and Ban et al. \cite{2005_BanAP_Rb2_destr}, was modeled as
   \begin{equation}
      4\,^{2}{\rm P}_{J} + {\rm K}_2  \overset{\sigma v}{\rightarrow} 3 \times 4\,^{2}{\rm S}_{1/2} \mbox{; $\sigma=10^{-14}$cm$^2$}
   \end{equation}
   \begin{equation}
      4\,^{2}{\rm P}_{J} + {\rm K}_2^+ \overset{\sigma v}{\rightarrow} 2 \times 4\,^{2}{\rm S}_{1/2} + {\rm K}^+ \mbox{; $\sigma=10^{-13}$cm$^2$}
   \end{equation}

   Reverse rates are calculated by detailed balance, where Tango et al.\cite{TangoEt_1968_K2Props}, and Magnier et al.\cite{2004_MagnierEt_K2Vs} provide details necessary to compute the molecular partition functions, as well as heat taken from, or put into, the thermal bath for the participants on each side to be fully thermalized.

   \section{Electron/neutral-mediated recombination (EMR/NMR)}
   \label{sec:A_3BR}

   As discussed in \cite{ZeldovichRaizerBook}, three-body recombination involves the probability an electron is close enough to an ion to be captured while colliding with a third body to remove energy and momentum:
   \begin{equation}
   \label{eqn:rec1}
      s_{\mboxss{3BR}} \sim \left(n_e \tfrac{4\pi}{3} r_c^3\right) \left(n_+ n_3 \langle (\pi r_t^2) v_{\mboxss{rel}} \rangle \right)   
   \end{equation}
   where $r_c = e/\langle E_e \rangle$ or $e/kT$ for a thermal population (in cgs units), and $\pi r_t^2$ is the energy-momentum transfer cross section.  The $r_t$ equals $r_c$ if the third body is a thermal electron.  This simple formalism applies most literally in the case of NMR at low density where third body impacts almost immediately re-ionizing the atom are unlikely.  Otherwise, the same impact ionizations counteracting electron-mediated recombinations should counteract neutral-mediated ones.


   \section{Electron impact energy transfer (EIT)}
   \label{sec:A_EIT}

   Phelps et al. \cite{1979_PhelpsEtal_eIX_K4S} measured cross sections for many impact transitions from the ground state, and the usual Klein-Rosseland relation gives the corresponding de-excitation cross sections in terms of lower/upper level degeneracies $g_l/g_u$, and transition energy;
   \begin{equation}
      \sigma_{\mboxss{dex}}(E) = \dfrac{g_l}{g_u}\dfrac{E+{E_t}}{E} \sigma_{\mboxss{exc}}(E+{E_t}).
   \end{equation}
   Note that the data were not fine-structure specific, so we assumed the branching ratios were purely based on degeneracies.  Other intermultiplet transitions were estimated based on the model in Vriens \& Smeets \cite{1980_VriensSmeets_eIX_SemiAnalytical}.  While one-electron, intramultiplet impact transitions are dipole forbidden, two-electron processes - with a threshold - are not, and Moores \& Sheorey \cite{1982_MooresSheorey_eIX_K} calculated such a cross-section for K(4\,$^2$P$_J$) mixing.

   Regarding methane, Itikawa et al. \cite{2004_Itikawa_eIX_MethaneEthane} provide mode-specific cross sections for the first vibrational modes, which are significantly larger than cross sections to excite rotational modes\cite{2015_SongEtal_eIX_Methane}.  We do not track individual levels of methane, as they will remain Boltzmann-distributed far more easily than potassium, so we use this assumption, plus the mode information (energies, degeneracies) in Itikawa et al. to include de-excitation from thermally excited methane.

   \section{Electron impact ionization (EII)}
   \label{sec:A_EII}

   Runs discussed here used a total cross section with the incident energy scaling derived by Gryzinski \cite{1965_Gryzinski1, 2004_TsipinyukEt_EII}:
   \begin{equation}
      \sigma^{\mboxss{G}}_{\mboxss{EII}} = \pi \left\langle r \right\rangle^2 \dfrac{1}{x} \left(\dfrac{x-1}{x+1}\right)^{3/2} \left(1+\dfrac{2}{3}\left(1-\dfrac{1}{2x}\right)\log[2.7+\sqrt{x-1}]\right)
   \end{equation}
   where $x$ is the ratio of incident kinetic to ionization energy, $E_{\mboxss{inc}}/W_{i}$, and $\left\langle r \right\rangle^2$ the squared mean radius of the outermost electron(s).  

   To simplify matters, a similar probability distribution for the secondary electron's energy based on \cite{1983_YoshidaPP_eii_UseOpbDistr,1971_OpalPB_eiiEDistr} was used for whatever total cross section was chosen above - as opposed to a certain differential cross section -  
   \begin{equation}
   \label{eqn:OPB71eII_EsDistrForm}
      \mathcal{P}(E_s) = \dfrac{1/(\bar{E}\arctan\left((E_p-W_i)/2\right))}{1+(E_{s}/\bar{E})^2}
   \end{equation}
   where $E_s$, and $E_p$ are the secondary's and primary's (initial) electron energies, and $\bar{E}$ is a parameter in their model set to 0.8$W_i$ for results presented here.  Since a source bin can map to many target bins in EII, the adjustment for global energy conservation in this case is done by scaling and skewing the entire product distribution slightly, leaving a linear system with two constraints (number and energy conservation), and two unknown parameters to solve.

   \section{Dissociative recombination and excitation (DR, DE)}
   \label{sec:A_DRDE}

   DR cross sections tend to share a power-law ``envelope'' inversely proportional to energy, and values up to $10^{-15}$cm$^2$ at 0.1 eV, but they can also fluctuate wildly in strength, and product branching ratio (see e.g. Little et al.\cite{2014_LittleEtal_DR_N2}).  Some of the reduction that better reproduces data, may really reflect steep, sporadic drops instead.  Once more urgent issues are addressed, examining the effect of toy notches versus overall re-scaling may be worthwhile.

   The explicit formula for ``default'' DR was $\sigma(E) = 10^{-15} \text{cm$^2$ } (E/\text{eV})^{-1}$, while the reduced-recombination runs used 0.05 times this, and the product branching ratio was a constant 50/50 split between 6\,$^2$S$_J$ and the 4\,$^2$D$_J$ multiplet.

   The toy model we used for DE was another power-law with sudden activation at the dissociation energy of K$_2^+$, so $\sigma_{\mboxss{DE}} = 3\times10^{15}\text{cm$^2$} (E/\text{eV})^{-1} : E>0.827\text{eV}$.


   \section{Drag}
   \label{sec:A_drag}
   \label{sec:A_LAST}

   For clarity, we spell out how we discretize drag.  At every wall indexed $j$, between cells indexed $j-1$ and $j$, the source contributions to bin densities, and the buffer thermal energy pool are:
   \begin{equation}
      \Delta s_{j} = -f\left[\dfrac{E_j-\kboltz{}T/2}{\tau_j}\dfrac{n_*}{\Delta E_*}\right] - 2\left[\dfrac{E_j \kboltz{}T}{\tau_j}\dfrac{n_j/\Delta{}E_j-n_{j-1}/\Delta{}E_{j-1}}{E_{j+1}-E_{j-1}}\right]
   \end{equation}
   \begin{equation}
      \Delta s_{j-1} = f \left[\dfrac{E_j-\kboltz{}T/2}{\tau_j}\dfrac{n_*}{\Delta E_*}\right] + 2\left[\dfrac{E_j \kboltz{}T}{\tau_j}\dfrac{n_j/\Delta{}E_j-n_{j-1}/\Delta{}E_{j-1}}{E_{j+1}-E_{j-1}}\right]
   \end{equation}
   \begin{equation}
      \Delta s_{\mboxss{eb}} = \tfrac{1}{2}\left( E_{j+1}-E_{j-1} \right) \Delta{s_{j-1}} 
   \end{equation}
   where $f=1$, and $n_*=n_j$ for $E_j>\kboltz{}T/2$, and $f=-1$, $n_*=n_{j-1}$ for $E_j<\kboltz{}T/2$.  The local energy exchange rate with the buffer, $\tau_j^{-1} = (m_e/m_{\mboxss{He}}) n_{\mboxss{He}} \sigma_{\mboxss{mt}}(E_j) v_e(E_j)$, where $\sigma_{\mboxss{mt}}$ is the momentum transfer cross section, $n_{\mboxss{He}}$ the buffer density, $m_{\mboxss{He}}$ the buffer particle mass, $m_{e}$ the electron mass, and $v_e$ the electron velocity.

\bibliography{trimmed_bib}

\end{document}